\documentclass[a4paper,english]{iopart}
\usepackage[]{fontenc}
\usepackage[latin1]{inputenc}

\makeatletter
\usepackage{ae,aecompl}
\usepackage{hyperref}

\usepackage{babel}
\makeatother
\begin{document}
\newcommand{\kl}[1]{{\textstyle #1}}

\hfill{}DESY 06-191

\title{Gravity and Quantum Fields in Discrete Space-Times}

\author{Florian Bauer}

\address{Deutsches Elektronen-Synchrotron DESY, Hamburg, Germany}

\ead{florian.bauer@desy.de}

\begin{abstract}
In a 6D model, where the extra dimensions form a discretised curved
disk, we investigate the mass spectra and profiles of gravitons and
Dirac fermions. The discretisation is performed in detail leading
to a star-like geometry. In addition, we use the curvature of the
disk to obtain the mass scales of this model in a more flexible way.
We also discuss some applications of this setup like generating small
fermion masses.
\end{abstract}
% Gravity in more than four dimensions, 04.50.+h
% Field theories in dimensions other than four, 11.10.Kk
% Neutrinos mass and mixing, 14.60.Pq

\section{Introduction}

We study a six-dimensional (6D) space-time consisting of a flat four-dimensional
(4D) subspace and a disk of constant curvature for the extra dimensions.
Furthermore, we discretise the disk in a way that~$N$ equidistant
lattice sites are situated on the boundary and a single site in the
centre of the disk~(section~\ref{sec:Disk-Curved-Disk}). Such star-like
geometries have been proven useful in various contexts, see for example~\cite{hep-th-0608176,Tomas,Gerhart,Disk-Casimir-1,Star-0,Star-1,Star-2}.
In this work we investigate the mass spectra and profiles of 4D gravitons
and fermions in this setup (sections~\ref{sec:Disk-Graviton-Spectrum}
and~\ref{sec:Disk-Fermions}), which has some nice applications.
For instance, we discuss a possibility to hide extra dimensions similar
to~\cite{ED-Hiding}, and we generate small fermion masses via a
discrete version of the wavefunction suppression mechanism~\cite{WaveFuncSuppression}.

\section{Curved Disk Geometry\label{sec:Disk-Curved-Disk}}

We consider a 6D model where both extra dimensions form a discretised
curved disk (for example a part of a 2-sphere) while the 4D subspace
remains continuous. Before the discretisation the space-time is described
by the line element\begin{equation}
\textrm{d}s^{2}=g_{\mu\nu}(x^{M})\textrm{d}x^{\mu}\textrm{d}x^{\nu}-(1-er^{2})^{-1}\textrm{d}r^{2}-r^{2}\textrm{d}\varphi^{2},\label{eq:Disk-6D-metric}\end{equation}
where~$x^{\mu}$ and~$x^{M}$ denote 4D and 6D coordinates, respectively.
The position on the disk is fixed by the polar coordinate~$\varphi:=x^{6}\in[0,2\pi]$
and the radial coordinate~$r:=x^{5}\in[0,L]$ with~$L$ being the
coordinate radius of the disk. From~(\ref{eq:Disk-6D-metric}) we
read off the metric components~$g_{\mu\nu}(x^{M})$ of the 4D subspace
as well as~$g_{55}=-(1-er^{2})^{-1}$ and~$g_{66}=-r^{2}$. Note
that the parameter~$e$ controls the curvature of the disk. For~$e>0$
the disk is spherically curved, whereas~$e<0$ leads to a hyperbolic
disk, and $e=0$ corresponds to a flat disk. Following~\cite{hep-th-0608176,Dissertation}
we now decompose the 6D Einstein-Hilbert action \begin{equation}
S=M_{6}^{4}\kl{\int}\textrm{d}^{6}x\,\sqrt{|g|}\, R=S_{{\rm 4D}}+S_{{\rm surface}}+S_{{\rm mass}},\label{eq:Disk-6D-EH-matter}\end{equation}
 into three parts, where $R$,~$g$ and~$M_{6}$ denote the 6D curvature
scalar, the determinant of the metric~$g_{MN}$ and the 6D Planck
scale, respectively. We find that the 4D curvature scalar~$R_{{\rm 4D}}$
in $S_{{\rm 4D}}:=M_{6}^{4}\int\textrm{d}^{6}x\sqrt{|g|}\,(R_{{\rm 4D}}+2e)$
contains only the 4D metric~$g_{\mu\nu}$ and derivatives with respect
to~$x^{\mu}$. The surface terms in~$S_{{\rm 4D}}$ vanish by choosing
suitable boundary conditions on the disk, and~$S_{{\rm mass}}$ is
given by\begin{equation}
S_{{\rm mass}}=M_{6}^{4}\int\textrm{d}^{6}x\sqrt{|g|}\sum_{c=5,6}\Big[-\frac{1}{4}g^{cc}g_{\mu\nu,c}(g^{\mu\nu}g^{\alpha\beta}-g^{\mu\alpha}g^{\nu\beta})g_{\alpha\beta,c}\Big].\end{equation}

Let us now introduce 4D graviton fields~$h_{\mu\nu}$ on a flat Minkowski
metric~$\eta_{\mu\nu}={\rm diag}(1,-1,-1,-1)$ by the expansion $g_{\mu\nu}\rightarrow\eta_{\mu\nu}+h_{\mu\nu}$.
As in~\cite{GravExDim} we choose a gauge, where we ignore graviphoton
and radion excitations, which could result from the~$g_{5M}$ and~$g_{6M}$
components of the metric. Since~$\eta_{\mu\nu}$ is constant we have~$g_{\mu\nu,A}\rightarrow h_{\mu\nu,A}$
and by expanding~$S_{{\rm mass}}$ in second order in~$h_{\mu\nu}$
we find\begin{eqnarray}
S_{{\rm mass}} & \rightarrow & M_{6}^{4}\int\textrm{d}^{4}x\,\textrm{d}\varphi\,\textrm{d}r\,\Big[+\frac{1}{4}\sqrt{\frac{g_{66}}{g_{55}}}\partial_{r}h_{\mu\nu}(\eta^{\mu\nu}\eta^{\alpha\beta}-\eta^{\mu\alpha}\eta^{\nu\beta})\partial_{r}h_{\alpha\beta}\nonumber \\
 &  & +\frac{1}{4}\sqrt{\frac{g_{55}}{g_{66}}}\partial_{\varphi}h_{\mu\nu}(\eta^{\mu\nu}\eta^{\alpha\beta}-\eta^{\mu\alpha}\eta^{\nu\beta})\partial_{\varphi}h_{\alpha\beta}\Big].\end{eqnarray}

Next, we discretise the disk by putting~$N$ lattice sites on the
boundary and a single site in centre of the disk so that only two
points are lying in radial direction. The coordinate distance between
the centre and each point on the boundary is given by the radius~$L$
of the disk, which in general differs from its proper radius. On the
boundary, the graviton fields are denoted by~$h_{\mu\nu}^{i}$ with~$i=1\dots N$,
and the position is given by~$\varphi^{i}=i\cdot\Delta\varphi$,
where $\Delta\varphi=2\pi/N$ is the angular lattice spacing. The
graviton field~$h_{\mu\nu}^{0}$ in the centre carries the index~$0$,
and the lattice spacing in radial direction is just~$\Delta r=L$.
Formally, we apply the following discretisation prescription to~$S_{{\rm mass}}$\[
\partial_{r}h(\varphi^{i})\rightarrow\frac{(h^{i}-h^{0})}{\Delta r},\qquad\partial_{\varphi}h(\varphi^{i})\rightarrow\frac{(h^{i+1}-h^{i})}{\Delta\varphi},\]
\begin{equation}
\int\textrm{d}r\, f(r)\rightarrow\Delta r\cdot f(L),\qquad\int\textrm{d}\varphi\, f(\varphi)\rightarrow\sum_{i=1}^{N}\Delta\varphi\cdot f(\varphi^{i}),\label{eq:Disk-DiscretPrescription}\end{equation}
where the integral~$\int\textrm{d}r$ is replaced by just one summation
interval of length~$L$ and the summand is evaluated at the position~$r=L$,
which avoids problems with the derivative~$\partial_{\varphi}$ at~$r=0$.
Thus we obtain (still non-diagonal) Fierz-Pauli mass terms~\cite{Fierz-Pauli}
for the gravitons on the discretised disk, that read (with~$h_{\mu\nu}^{N+1}\equiv h_{\mu\nu}^{1}$)\begin{eqnarray}
\fl{}S_{{\rm mass}}\rightarrow M_{4}^{2}\int\textrm{d}^{4}x\sum_{i=1}^{N}\big[m_{\star}^{2}\cdot(h_{\mu\nu}^{i}-h_{\mu\nu}^{0})(\eta^{\mu\nu}\eta^{\alpha\beta}-\eta^{\mu\alpha}\eta^{\nu\beta})(h_{\alpha\beta}^{i}-h_{\alpha\beta}^{0})\nonumber \\
\fl{}\qquad\qquad\qquad\qquad+m^{2}\cdot(h_{\mu\nu}^{i+1}-h_{\mu\nu}^{i})(\eta^{\mu\nu}\eta^{\alpha\beta}-\eta^{\mu\alpha}\eta^{\nu\beta})(h_{\alpha\beta}^{i+1}-h_{\alpha\beta}^{i})\big].\label{eq:Disk-S-graviton-mass}\end{eqnarray}
Note that the actual graviton mass scale from the radial derivatives,~$m_{\star}$,
and respectively from the angular derivatives,~$m$, depend on the
4D Planck mass~$M_{4}$ of the observer's site (brane):\begin{equation}
m_{\star}^{2}:=\frac{M_{6}^{4}}{{4M}_{4}^{2}}\cdot\frac{2\pi\sqrt{1-eL^{2}}}{N},\qquad m^{2}:=\frac{M_{6}^{4}}{{4M}_{4}^{2}}\cdot\frac{N}{2\pi\sqrt{1-eL^{2}}}.\label{eq:Disk-mstar-m}\end{equation}
However, the ratio of masses is independent of the Planck scales,\begin{equation}
\frac{m_{\star}^{2}}{m^{2}}=\frac{(2\pi)^{2}}{N^{2}}(1-eL^{2}),\label{eq:Disk-Graviton-Mass-Ratio}\end{equation}
which also shows that arbitrarily large hierarchies between~$m_{\star}$
and~$m$ are possible by choosing the disk parameters~$eL^{2}$
and~$N$ appropriately.

In order to determine the 4D Planck mass~$M_{4}$ on the sites we
need the proper area of the curved disk, which is given by\begin{equation}
A:=\int_{0}^{2\pi}\textrm{d}\varphi\int_{0}^{L}\textrm{d}r\sqrt{|g_{55}g_{66}|}=\frac{2\pi}{e}(1-\sqrt{1-eL^{2}}).\label{eq:Disk-proper-area}\end{equation}

We now proceed to discretise~$S_{{\rm 4D}}$. Since the extra-dimensional
disk has a constant curvature it is well motivated that~$M_{4}$
should be constant and universal on all sites, too. Thus the Einstein-Hilbert
terms of all~$N+1$ sites must have the form\begin{equation}
M_{4}^{2}\sum_{i=0}^{N}\int\textrm{d}^{4}x\, R_{{\rm 4D}},\label{eq:Disk-Sum-EH}\end{equation}
which does not depend on~$r$ and~$\varphi$ anymore. By comparing
this term with~$S_{{\rm 4D}}=M_{6}^{4}\int\textrm{d}^{6}x\sqrt{|g|}R_{{\rm 4D}}$
we find that the 4D Planck scale on the sites is fixed by $M_{4}^{2}=M_{6}^{4}A/(N+1)$,
where we used~(\ref{eq:Disk-proper-area}) in~$S_{{\rm 4D}}$ and
evaluated the sum in~(\ref{eq:Disk-Sum-EH}). However, we remark
that~$M_{4}$ is not the (reduced) Planck scale $M_{{\rm Pl}}=1/(8\pi G)\sim10^{18}\:\textrm{GeV}$
that couples gravity to 4D matter. But~$M_{{\rm Pl}}$ is determined
by integrating out the extra dimensions in the continuum, which means
$M_{6}^{4}\int\textrm{d}^{6}x\:\sqrt{|g|}R_{{\rm 4D}}=M_{{\rm Pl}}^{2}\int\textrm{d}^{4}x\: R_{{\rm 4D}}$
and thus $M_{{\rm Pl}}^{2}=M_{6}^{4}A=(N+1)M_{4}^{2}$.

\section{Graviton Mass Spectrum\label{sec:Disk-Graviton-Spectrum}}

We omit to show the kinetic terms for the 4D gravitons~$h_{\mu\nu}^{i}$.
They just follow from applying the graviton expansion to the Einstein-Hilbert
terms in~(\ref{eq:Disk-Sum-EH}), see e.g.~\cite{Graviton-Kinetic}.
To determine the graviton mass spectrum we have to diagonalise~$S_{{\rm mass}}$
in~(\ref{eq:Disk-S-graviton-mass}) by a unitary transformation.
If we denote the graviton mass eigenstates by~$H_{\mu\nu}^{n}$,
corresponding to the masses~$M_{n}$, we find the following relations
for the eigenvectors:\begin{eqnarray}
H_{\mu\nu}^{0} & = & \frac{1}{\sqrt{N+1}}\sum_{i=0}^{N}h_{\mu\nu}^{i},\label{eq:Disk-EV-H0}\\
H_{\mu\nu}^{p} & = & \frac{1}{\sqrt{N}}\sum_{i=1}^{N}\left[\sin(2\pi\kl{\frac{p}{N}})+\cos(2\pi\kl{\frac{p}{N}})\right]\cdot h_{\mu\nu}^{i},\label{eq:Disk-EV-Hp}\\
H_{\mu\nu}^{N} & = & \frac{1}{\sqrt{N(N+1)}}\left[-N\cdot h_{\mu\nu}^{0}+\sum_{i=1}^{N}h_{\mu\nu}^{i}\right],\label{eq:Disk-EV-HN}\end{eqnarray}
where~$p=1,\dots,N-1$. The eigenvalues~$M_{n}$ are respectively
given by\begin{equation}
M_{0}^{2}=0,\qquad M_{p}^{2}=m_{\star}^{2}+4m^{2}\textrm{sin}^{2}\frac{\pi p}{N},\qquad M_{N}^{2}=(N+1)m_{\star}^{2}.\label{eq:Disk-Graviton-Masses}\end{equation}
From these results we observe that the zero-mode~$H_{\mu\nu}^{0}$
has a flat profile and is equally located on all sites, whereas the
mode~$H_{\mu\nu}^{N}$ with squared mass~$(N+1)m_{\star}^{2}$ is
peaked on the centre site with equal support on the boundary sites.
The modes~$H_{\mu\nu}^{p}$ with~$p=1,\dots,N-1$ are located only
on the boundary with a typical finite Kaluza-Klein (KK) mass spectrum
that has been shifted by~$m_{\star}^{2}$. In the limit~$m\ll m_{\star}$
the masses of the states~$H_{\mu\nu}^{p}$ in~(\ref{eq:Disk-EV-Hp})
become degenerate, and for~$N\gg1$ the mode~$H_{\mu\nu}^{N}$ becomes
very heavy. Note that the latter case can be realised by a sufficiently
large negative curvature of the disk, which is a clear advantage over
a flat disk model.

Finally, we mention that a scenario related to ours has been discussed
recently in the context of multi-throat geometries~\cite{ED-Hiding}.
It was shown that large extra dimensions can be hidden in the sense
that the occurrence of massive KK modes is shifted to energies much
higher than the compactification scale of the extra dimension, which
helps evading limits on KK particles. In our model this behaviour
can be observed for the modes~$H_{\mu\nu}^{n>0}$ in the limit~$m_{\star}\gg m$,
too.

\section{Fermions on the Disk\label{sec:Disk-Fermions}}

Let us now investigate the incorporation of Dirac fermions into the
discretised disk model of section~\ref{sec:Disk-Curved-Disk}. As
for the graviton case we start with a 6D Dirac fermion~$\Psi$ in
the continuum. Using the vielbein formalism, the corresponding action~$S$
on the curved disk reads\begin{equation}
S=\int\textrm{d}^{6}x\,\sqrt{|g|}\left[\frac{1}{2}\rmi\left(\overline{\Psi}G^{A}V_{A}^{M}\nabla_{M}\Psi-\overline{\nabla_{M}\Psi}V_{A}^{M}G^{A}\Psi\right)\right]\!,\label{eq:Disk-6D-Fermion-Action}\end{equation}
where we denote 6D Lorentz indices by~$A,B,\dots$ and general coordinate
indices by~$M,N,\dots$, respectively. Moreover,~$G^{A}$ are 6D
Dirac matrices, and for the barred spinor~$\overline{\Psi}$ we use
the abbreviation $\overline{\Psi}=\Psi^{\dag}G^{0}$. The vielbein
components~$V_{A}^{M}(x^{N})$ follow from the relation~$g^{MN}=V_{A}^{M}V_{B}^{N}\eta^{AB}$,
which connects the Lorentz coordinate system with the general coordinate
system. For the diagonal metric~(\ref{eq:Disk-6D-metric}), we find~$V_{A}^{M}=\delta_{A}^{M}$
with the exceptions~$V_{A=5}^{M=5}=\sqrt{|g^{55}|}=:V_{5}$ and~$V_{A=6}^{M=6}=\sqrt{|g^{66}|}=:V_{6}$.
On a curved space-time the covariant derivative~$\nabla_{M}=\partial_{M}+\Gamma_{M}$
for spinors contains in addition to the usual partial derivative~$\partial_{M}$
also the spin connection~$\Gamma_{M}=\frac{1}{8}[G^{A},G^{B}]V_{A}^{N}V_{BN;M}.$ 

To determine the form of the 6D $\gamma$-matrices~\cite{6D-Fermions}
let us first look at the 4D case, where the $\gamma$-matrices are
given by \begin{equation}
\gamma^{0}=\left[\begin{array}{cc}
0 & 1_{2}\\
1_{2} & 0\end{array}\right]\!,\qquad\gamma^{k}=\left[\begin{array}{cc}
0 & \sigma^{k}\\
-\sigma^{k} & 0\end{array}\right]\!,\qquad\gamma^{5}=\rmi\gamma^{0}\gamma^{1}\gamma^{2}\gamma^{3}.\end{equation}
Here,~$k=1\dots3$ and~$\sigma^{k}$ denote the Pauli matrices.
In five dimensions the number of spinor components is still four and
the corresponding $\gamma$-matrices are simply given by~$\Gamma^{0}=\gamma^{0}$,
$\Gamma^{k}=\gamma^{k}$ and~$\Gamma^{5}=\rmi\gamma^{5}=-(\Gamma^{5})^{\dag}$.
In six dimensions, however, the Dirac algebra is 8-dimensional, where
we use the following set of $\gamma$-matrices\begin{eqnarray}
G^{0} & = & \left[\begin{array}{cc}
0 & 1_{4}\\
1_{4} & 0\end{array}\right]\!,\qquad G^{6}=\left[\begin{array}{cc}
0 & \Gamma^{0}\\
-\Gamma^{0} & 0\end{array}\right]=-(G^{6})^{\dag},\\
G^{n} & = & \left[\begin{array}{cc}
0 & \Gamma^{0}\Gamma^{n}\\
-\Gamma^{0}\Gamma^{n} & 0\end{array}\right]=-(G^{n})^{\dag},\qquad n=1,2,3,5,\end{eqnarray}
which fulfil the Cifford algebra~$\{ G^{A},G^{B}\}=2\eta^{AB}\cdot1_{8}$. 

From the form of the vielbeins and the $\gamma$-matrices it follows
that all spin connection components~$\Gamma_{M}$ vanish except for
$\Gamma_{6}=\frac{1}{4}[G^{5},G^{6}]\sqrt{1-er^{2}}$. And because
of~$\Gamma_{6}=-\Gamma_{6}^{\dag}$ the terms involving~$\Gamma_{6}$
in~(\ref{eq:Disk-6D-Fermion-Action}) cancel.

Let us now diagonalise the fermion action by the substitution~$\Psi=G^{6}\Phi$.
Then the decomposition of the eight-component spinor~$\Phi=(\Phi_{a},\Phi_{b})^{\textrm{T}}$
into two four-component spinors~$\Phi_{a}$, $\Phi_{b}$ yields in~(\ref{eq:Disk-6D-Fermion-Action})
\begin{eqnarray}
\rmi\overline{\Psi}G^{A}V_{A}^{M}\nabla_{M}\Psi=\rmi\left[\overline{\Phi_{a}},\overline{\Phi_{b}}\right]\times\left[\left(\begin{array}{cc}
\gamma^{0} & 0\\
0 & \gamma^{0}\end{array}\right)\partial_{0}+\left(\begin{array}{cc}
-\gamma^{k} & 0\\
0 & \gamma^{k}\end{array}\right)\partial_{k}\right.\nonumber \\
\qquad+\left.\left(\begin{array}{cc}
-\rmi\gamma^{5} & 0\\
0 & +\rmi\gamma^{5}\end{array}\right)V_{5}\partial_{5}+\left(\begin{array}{cc}
1 & 0\\
0 & -1\end{array}\right)V_{6}\partial_{6}\right]\times\left[\begin{array}{c}
\Phi_{a}\\
\Phi_{b}\end{array}\right]\end{eqnarray}
with~$\overline{\Phi_{a,b}}=\Phi_{a,b}^{\dag}\gamma^{0}$. From the
last line one can read off that~$\Phi_{a}$ corresponds to~$\Phi_{b}$
but with negative energy, therefore we will work only with~$\Phi_{b}$
in the following. If we now denote the left- and right-handed components
of~$\Phi_{b}$ by~$\Phi_{{\rm L,R}}:=\frac{1}{2}(1\mp\gamma^{5})\Phi_{b}$,
then the full action for~$\Phi_{b}$ can be written in the form\begin{eqnarray}
S & = & \int\textrm{d}^{6}x\sqrt{|g|}\Big[\frac{1}{2}\rmi\left(\overline{\Phi_{b}}\gamma^{\mu}\partial_{\mu}\Phi_{b}-\overline{\partial_{\mu}\Phi_{b}}\gamma^{\mu}\Phi_{b}\right)\nonumber \\
 & - & V_{5}\frac{1}{2}\left(\overline{\Phi_{{\rm L}}}\partial_{5}\Phi_{{\rm R}}+\overline{\partial_{5}\Phi_{{\rm R}}}\Phi_{{\rm L}}\right)-\rmi V_{6}\frac{1}{2}\left(\overline{\Phi_{{\rm L}}}\partial_{6}\Phi_{{\rm R}}-\overline{\partial_{6}\Phi_{{\rm R}}}\Phi_{{\rm L}}\right)\Big],\end{eqnarray}
where we have applied the boundary conditions of section~\ref{sec:Disk-Curved-Disk}
after integration by parts.

Since 6D spinors have mass dimension~$\frac{5}{2}$ we have to rescale
them in order to obtain usual 4D spinors. As in the graviton case
we integrate the kinetic terms over the extra dimensions and apply
a similar discretisation procedure as in section~\ref{sec:Disk-Curved-Disk},
which means\begin{equation}
\int\textrm{d}^{6}x\sqrt{|g|}\frac{1}{2}\rmi\overline{\Phi_{b}}\gamma^{\mu}\partial_{\mu}\Phi_{b}\rightarrow\sum_{j=0}^{N}\frac{A}{N+1}\int\textrm{d}^{4}x\frac{1}{2}\rmi\overline{\Phi_{b}^{j}}\gamma^{\mu}\partial_{\mu}\Phi_{b}^{j},\end{equation}
where~$A$ is the proper area given in~(\ref{eq:Disk-proper-area}).
Finally, we absorb the factor~$A/(N+1)$ into the fermion fields
$\chi:=\Phi_{b}\sqrt{A/(N+1)}$ and subsequently apply the discretisation
prescriptions~(\ref{eq:Disk-DiscretPrescription}) with~$h_{\mu\nu}$
replaced by~$\chi$. As a result we obtain the action for~$N+1$
4D fermions, ($\chi^{N+1}\equiv\chi^{1}$)\begin{eqnarray}
S & = & \sum_{j=0}^{N}\int\textrm{d}^{4}x\frac{1}{2}\rmi\left(\overline{\chi^{j}}\gamma^{\mu}\partial_{\mu}\chi^{j}-\overline{\partial_{\mu}\chi^{j}}\gamma^{\mu}\chi^{j}\right)\nonumber \\
 & - & \sum_{j=1}^{N}\int\textrm{d}^{4}x\cdot m_{\star}\left(\overline{\chi_{{\rm L}}^{j}}(\chi_{{\rm R}}^{j}-\chi_{{\rm R}}^{0})+(\overline{\chi_{{\rm R}}^{j}}-\overline{\chi_{{\rm R}}^{0}})\chi_{{\rm L}}^{j}\right)\nonumber \\
 & - & \sum_{j=1}^{N}\int\textrm{d}^{4}x\cdot\rmi m\left(\overline{\chi_{{\rm L}}^{j}}(\chi_{{\rm R}}^{j+1}-\chi_{{\rm R}}^{j})-(\overline{\chi_{{\rm R}}^{j+1}}-\overline{\chi_{{\rm R}}^{j}})\chi_{{\rm L}}^{j}\right)\end{eqnarray}
with the mass scales~$m_{\star}:=2\pi L(N+1)/(AN)$ and~$m:=L(N+1)/(A\sqrt{1-eL^{2}})$.
Hence, the ratio~$m_{\star}^{2}/m^{2}$ is the same ratio as in~(\ref{eq:Disk-Graviton-Mass-Ratio})
for the gravitons. Next, we apply a bi-unitary transformation relating
the states~$\chi$ to the mass eigenstates~$\psi$:\begin{eqnarray}
\overline{\chi_{{\rm L}}^{0}} & = & \overline{\psi_{{\rm L}}^{0}},\qquad\overline{\chi_{{\rm L}}^{j}}=\frac{1}{\sqrt{N}}\sum_{n=1}^{N}\exp(+2\pi\rmi\cdot j\kl{\frac{n}{N}})\overline{\psi_{{\rm L}}^{n}},\nonumber \\
\chi_{{\rm R}}^{0} & = & \frac{1}{\sqrt{N+1}}\psi_{{\rm R}}^{0}-\frac{N}{\sqrt{N(N+1)}}\psi_{{\rm R}}^{N},\label{eq:Disk-Fermion-Trafo}\\
\chi_{{\rm R}}^{j} & = & \frac{1}{\sqrt{N}}\sum_{n=1}^{N-1}\exp(-2\pi\rmi\cdot j\kl{\frac{n}{N}})\psi_{{\rm R}}^{n}+\frac{1}{\sqrt{N+1}}\psi_{{\rm R}}^{0}+\frac{1}{\sqrt{N(N+1)}}\psi_{{\rm R}}^{N}.\nonumber \end{eqnarray}
The corresponding mass spectrum contains one massless fermion~$\psi^{0}$,
one heavy fermion~$\psi^{N}$ with mass~$m_{\star}\sqrt{N+1}$ and~$N-1$
fermions~$\psi^{1},\dots,\psi^{N-1}$ with squared absolute mass
values~$m_{\star}^{2}+4m^{2}\sin^{2}(\frac{\pi n}{N})+2m_{\star}m\sin(\frac{2\pi n}{N})$.
In contrast to the graviton mass spectrum~(\ref{eq:Disk-Graviton-Masses}),
here we find an additional interference term~$\propto m_{\star}m$,
which can be removed by a slightly modified discretisation procedure
for the angular direction. Instead of~$\partial_{6}\chi\rightarrow(\chi^{j+1}-\chi^{j})/\Delta\varphi$
we use the prescription~$\partial_{6}\chi\rightarrow\rmi(\chi^{j+\frac{1}{2}}-\chi^{j-\frac{1}{2}})/\Delta\varphi$.
This does not change the zero more or the heavy mode, but the transformations
in~(\ref{eq:Disk-Fermion-Trafo}) lead now to the mass spectrum~$m_{\star}^{2}+4m^{2}\sin^{2}(\frac{\pi n}{N})$
for the modes~$\psi^{1}\dots\psi^{N-1}$, which has exactly the same
structure as that of the gravitons in~(\ref{eq:Disk-Graviton-Masses}). 

Our results for the fermions on the discretised disk can be applied
directly to generate small fermion masses. For this purpose we put
the standard model of particles~(SM) on the centre site of our disk.
In this place the left-handed SM lepton doublet~$\ell$ may couple
to the 4D component~$\chi_{{\rm R}}^{0}$ of the 6D Dirac field and
to the vacuum expectation value~$\langle H\rangle$ of the Higgs
doublet via an Yukawa interaction term schematically given by~$\overline{\ell}\langle H\rangle\chi_{{\rm R}}^{0}$.
Now, a large number~$N\sim10^{24}$ of lattice sites lets~$\Phi^{N}$
decouple due to its large mass~$m_{\star}\sqrt{N+1}$, and~(\ref{eq:Disk-Fermion-Trafo})
shows that the right-handed fermion~$\chi_{{\rm R}}^{0}$ on the
centre site essentially consists only of the zero-mode~$\psi_{{\rm R}}^{0}$
with a tiny weight factor~$1/\sqrt{N+1}$. Thus the Yukawa interaction
of~$\ell$ with~$\chi_{{\rm R}}^{0}$, \begin{equation}
\overline{\ell}\langle H\rangle\chi_{{\rm R}}^{0}\sim\frac{1}{\sqrt{N+1}}\nu_{{\rm L}}\langle H\rangle\psi_{{\rm R}}^{0},\end{equation}
leads to a strong suppression of the SM neutrino~($\nu_{{\rm L}}$)
mass, representing a discrete version~\cite{hep-th-0608176} of the
wave function suppression mechanism in continuous higher dimensions~\cite{WaveFuncSuppression}.

\section{Conclusions\label{sec:Disk-Conclusions}}

Our 6D model with a discretised extra-dimensional curved disk leads
to mass spectra that have the same structure for gravitons and fermions.
Moreover, the special discretisation of the disk allows that the ratio
of mass scales in the spectra can be adjusted in a flexible manner
by the parameters of the disk. It is thus possible to obtain a gap
between the zero mode and the first massive mode that is much larger
than the gap between the other massive modes. We have also discussed
the generation of small SM fermion masses in this setup. Finally,
we mention that the strong coupling regime of this model and a more
refined scenario including warping were investigated in~\cite{hep-th-0608176},
where some of our results have been applied, too.

\end{document}